# Species Diversity in Rock-Paper-Scissors Game Coupling with Levy Flight

Dong Wang[1,2], Qian Zhuang[1], Jing Zhang[1], Zengru Di[1*],

*1. Department of Basic Science, Naval Aeronautical And Astronautical University, Yantai 264001, China*

*2. Department of Systems Science, School of Management and Center for Complexity Research, Beijing Normal University, Beijing 100875, China*

## Abstract

Rock–paper–scissors (RPS) game is a nice model to study the biodiversity in ecosystem. However, the previous studies only consider the nearest- neighbor-interaction among the species. In this paper, taking the long range migration into account, the effects of the interplay between nearest-neighbor-interaction and long-range-interaction of Levy flight with distance distribution $l^h$ (-0.3<$h$<-0.1) in spatial RPS game is investigated. Taking the probability of long range Levy flight and the power exponent as parameters, the coexistence conditions of three species are found. The critical curves for stable coexistence of three species in the parameters space are presented. It is also found that long-range-interaction with Levy flight has interesting effects on the final spatiotemporal pattern of the system. The results reveal that the long-range-interaction of Levy flight exhibit pronounced effects on biodiversity of ecosystem.

## 1. Introduction

Rock–paper–scissors (RPS) game describes a basic kind of cyclic, non-transitive interaction among competing populations [1-4]. Experimental and theoretical investigations have shown that diversity of some species in ecosystem is maintained by this type of mechanism.

The previous theoretical works are mainly based on nearest-neighbor-interactions (NNI) [5-10]. Recently RPS game, provided with the interplay between the interaction range and mobility on coexistence, was studied [11, 12]. In reality, individuals in populations may go long distance before predation or reproduction. For example, birds fly through long distance apart from its' habitat and prey, in special seasons some kinds of birds migrate over very long distance, other animals do so in different territories in some season, bacteria swim and tumble in the air or water, and so on. So long-range-interaction (LRI) is universal in the process of biological evolution.

---

* Author for Correspondence: zdi@bnu.edu.cn

Empirical studies on the pattern of many living organisms' moving have revealed the existence of Levy flight. Now observations of Levy flight have been extended to many species [13-21]. Levy flight is characterized by many small steps connected by longer relocations; the same sites are revisited scarcely. The probability density function for the traveling distance has a power-law tail in the long-distance regime, as fellow

$$P(l_i) \sim l_i^h \qquad (-3 \leq h < -1),$$

$l_i$ is the flight length of step $i$, $h$ is the power-law exponent. Levy-flight foraging hypothesis [22-24] predicts that Levy flight have higher search efficiencies in environments where prey is scarce, nevertheless a Brownian walk take place more likely where prey is abundant. Hence there exist NNI and LRI among species, such as selection, reproduction and migration. Based on above ideas, we explore the effects of long-range-interactions upon species diversity in the framework of RPS game in spatially extended ecological system in this paper.

The model is described in Section 2. Section 3 gives some simulation results. Taking the probability of long range Levy flight and the power exponent as parameters, the coexistence conditions of three species are found. The critical curves for stable coexistence of three species in the parameters space are presented. They display that the higher the probability of long range interaction is and the larger the exponent $h$ is, the easier the coexistence state of three species is destroyed. More interestingly, compared with the only nearest–neighbor -interaction, a small fraction of long-range-interaction with Levy flight generate more fragments in the patterns, lead to spatial heterogeneity and patchiness which are common in ecosystem. However, the final steady pattern forms larger dense spiral wave when there is only long-range-interaction. The results reveal that the long-range-interaction of Levy flight exhibit pronounced effects on biodiversity of ecosystem.

## 2. The spatial RPS game with Levy flight

Consider a spatial and stochastic model of cyclically competing populations. Three species $A$, $B$ and $C$ populate a square lattice of $N \times N$ sites with periodic boundary condition. Each site is occupied by one individual or left empty, $\phi$ denotes empty site. Agent can interact with each other by cyclic competition,

following the rules

$$AB \xrightarrow{\mu} A\phi, BC \xrightarrow{\mu} B\phi, CA \xrightarrow{\mu} C\phi, \quad (1)$$

$$A\phi \xrightarrow{\delta} AA, B\phi \xrightarrow{\delta} BB, C\phi \xrightarrow{\delta} CC, \quad (2)$$

$$XY \xrightarrow{e} YX, \quad (X,Y \in \{A,B,C,\phi\}) \quad (3)$$

Equations (1) reflect cyclic selection: *A* can kill or prey *B*, yielding an empty site. In the same way, *B* outperforms *C*, and *C* in turn defeats *A*. Selection occurs at rate $\mu$. Equations (2) describe reproduction of individuals at rate $\delta$, which is only allowed on empty site. In Equation (3), individual is able to swap position with each other at exchange rate $e$, including empty sites.

Our interest is focused on the interplay between NNI and LRI in spatial RPS game. We choose Levy flight as the pattern of LRI. In stochastic lattice simulations with periodic boundary conditions, the initial densities of three species and empty sites are always fixed as 25％ equally. At each Monte Carlo simulation step, firstly a random individual is chosen，then the interaction pattern is decided randomly. NNI is chosen with probability $p_0$ and LRI of Levy flight is chosen with probability 1-$p_0$. When the interaction is NNI, the individual previously chosen will interact with one of its four nearest neighbors, which is also randomly determined. When the interaction is LRI, the individual previously chosen will interact with another site which is decided by Levy flight using the algorithm of Monte Carlo simulation. We first chose a distance *l* (manhatton distance in this paper) from the power law distribution $P(l_i) \sim l_i^h$, and then randomly chose a site from all the sites with the same distance. Then, according to the algorithm of Gillespie [25,26], the probabilities of the three possible reactions in equation (1)、(2) and (3), are computed by using the reaction rates. Selection occurs with the probability $\frac{\mu}{\mu+\delta+e}$, reproduction with the probability $\frac{\delta}{\mu+\delta+e}$, and exchange occurs with the probability $\frac{e}{\mu+\delta+e}$. When every individual has reacted on average once, one generation was set as the unit of time.

Taking account of species in realistic situation, individual walks or flies under the constraints of its bio-energy and territory. So we propose that each individual's step distributed from 1 to a maximal length (denoted by $l_{max}$) when Levy flights is chosen. Concisely selection rate $\mu$ and reproduction rate $\delta$ are set equal to 1 in the simulation. In our stochastic lattice simulations, we mainly studied coexistence condition of three species. We define the condition of species extinction so far as one of the three species extinct in the simulations.

## 3. The simulation results

### (1) Critical curves of $p_0$ vs. $h$ for given $e$ and $l_{max}$

We set the exchange rate $e = 0.2$, then study the coexistence condition of three species: the critical curves of parameter $p_0$ and power-law exponent $h$ with the maximal step length $l_{max}$ of Levy flight changes on the lattice with 100×100 sites.

In the stochastic lattice simulations, we employ $10^5$ generations of each realization. Fig. 1(a-b) show the results. Each curve means critical and stable coexistence of three species in Fig. 1(a-b).

Proposing else condition are unchanged. Likewise we find the LRI effects of Levy flight are pronounced. As the parameter $p_0$ is reduced (means the probability of choosing Levy flight increases), power-law exponent $h$ must decrease for stable coexistence of three species. In Fig.1(a), when maximal step length $l_{max}$ increases from 5 to 50, the area which represent stable coexistence of three species on the graph diminish gradually. The curve of $l_{max} = 50$ is closer to a margin. Along with the increment of maximal step length $l_{max}$ from 50 to 105 in Fig.1(b), we could find that the area of stable coexistence of three species vary a little. Furthermore the area of stable coexistence state tends to a minimal limit along with increment of the maximal step length $l_{max}$ in Fig. 1(a-b). These results show that the long-range interaction with Levy flight affect the stability of the coexistence state strikingly.

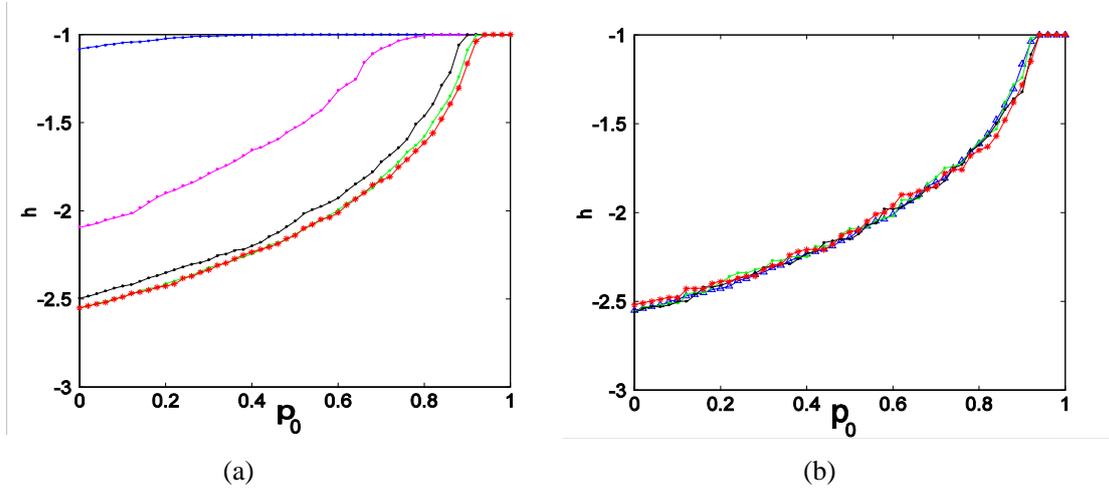

(a) (b)

Fig.1. Relation between parameter $p_0$ and power-law exponent $h$ in the context of different maximal step length $l_{max}$ of Levy flight. Colors represent different maximal step length of $l_{max}$, in (a): 5 (blue circle), 10 (magenta circle), 20 (black circle), 30 (green circle), 50 (red flake); in (b): 50 (blue triangle), 65 (green circle), 85 (black circle), 105 (red flake). The horizontal ordinate represents the parameter $p_0$, vertical coordinate represents power-law exponent $h$. Each point is averaged over 30 random realizations of stochastic simulations employing same initial densities on a $100 \times 100$ square lattice. Each realization consists of $10^5$ generations. Above each curve is extinction region of three species, correspondingly below (and including ) each curve is stable coexistence region of three species.

### (2) The results of the small system

In order to uncover whether lattice scale have effect on stable coexistence condition of three species, we reduce the square lattice into $50 \times 50$ for further simulations. The result is displayed in Fig.2. Each curve represents critical and stable coexistence of three species. The same as the above simulations, each realization is the results of $10^5$ generations. It could be found that the area of stable coexistence of three species become much lesser than that of $100 \times 100$ square lattice, although the trend of variation is the same. Similarly the effects of Levy flight are displayed obviously. When the parameter $p_0$ is reduced, power-law exponent $h$ must be decreased for stable coexistence of three species. With the increase of maximal step length $l_{max}$ from 2 to 25, the area representing stable coexistence of three species on the graph shrinks gradually. Furthermore in Fig.2, with the further increment of maximal step length $l_{max}$ from 25 to 55, the area of stable coexistence of three species only changes a little. It also appears that the area of stable coexistence have a minimal limit with the increase of maximal step length $l_{max}$. Compared with the results in the last subsection, we can see the size of the system affects the results obviously.

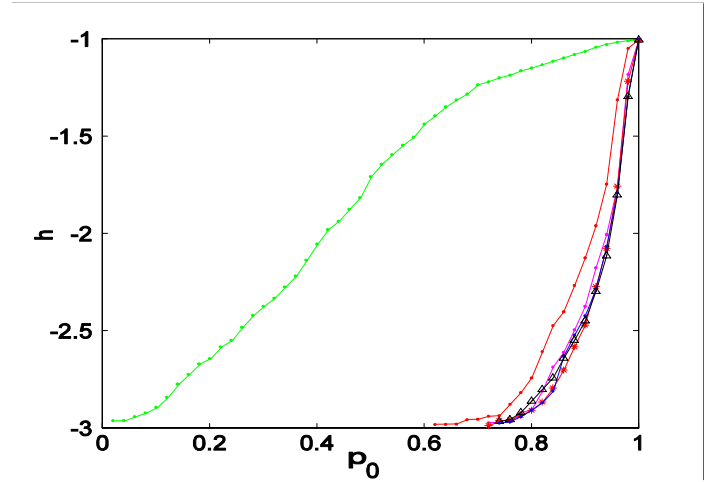

Fig.2. Critical curves of parameter $p_0$ and power-law exponent $h$ with different maximal step length $l_{max}$ of Levy flight. The square lattice is $50 \times 50$. Colors represent different maximal step length of $l_{max}$: 2(green circle), 10 (red circle), 20 (magenta circle), 25 (red flake), 30 (blue circle), 50 (black triangle). The horizontal and vertical coordinate represent the same parameters as Fig.1. Each point is averaged over 30 random realizations of stochastic simulations of same initial densities. Each realization contains $10^5$ generations.

### (3) The spatiotemporal pattern under Levy flight

In stochastic lattice simulations, we find the LRI of Levy flight also changes the spatial pattern in RPS game. Fig.3 and Fig.4 show snapshots of the patterns for different values of parameters, the lattice sizes are $100 \times 100$ and $500 \times 500$ respectively. Individuals of three competing species $A$ (red), $B$ (yellow), $C$ (light blue) and empty site (dark blue) occupy the sites of the lattice.

The patterns in Fig.3 and Fig.4 display some characters different from the previous work [5-12]. Furthermore, we can see that some individuals of three species move with Levy flight behavior clearly. On condition that other parameters are unchangeable. Firstly, as $p_0 = 1.0 \to 0.0$ (means that the probability of Levy flight increases), the number of species patches abruptly increase and then succession decrease gradually. Secondly, when $h = -1.0 \to -3.0$, the number of species patches increase inchmeal. Compared with NNI, the introduce of LRI of Levy flight generate more fragments in the spatial patterns initially, lead to spatial heterogeneity and patchiness. Those results are similar to the characters of species patches in ecosystem. As maximal step length $l_{max}$ increases, the number of species patches decrease, and

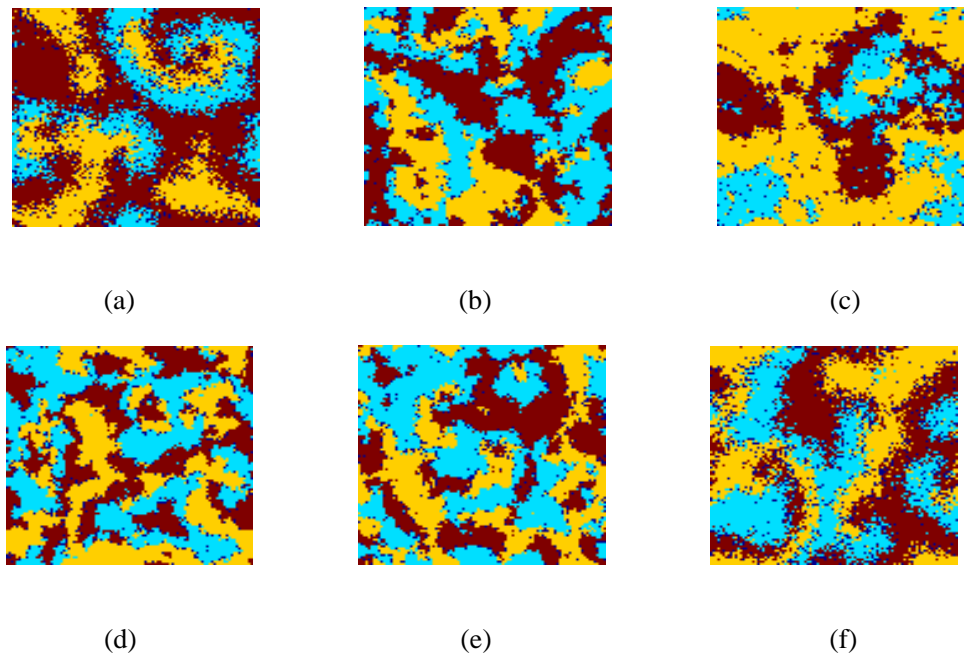

(a)  (b)  (c)

(d)  (e)  (f)

Fig.3. Snapshots of patterns. The size of square lattice is $100 \times 100$. For each realization, the simulation consists of $10^4$ time steps. The maximal step length $l_{max}$ of Levy flight equals 50.

(a) $e = 0.1, p_0 = 1.0$.   (b) $h = -2.0, e = 0.1, p_0 = 0.9$.
(c) $h = -2.0, e = 0.1, p_0 = 0.5$.   (d) $h = -3.0, e = 0.1, p_0 = 0.9$.
(e) $h = -3.0, e = 0.1, p_0 = 0.5$.   (f) $h = -3.0, e = 0.1, p_0 = 0.0$.

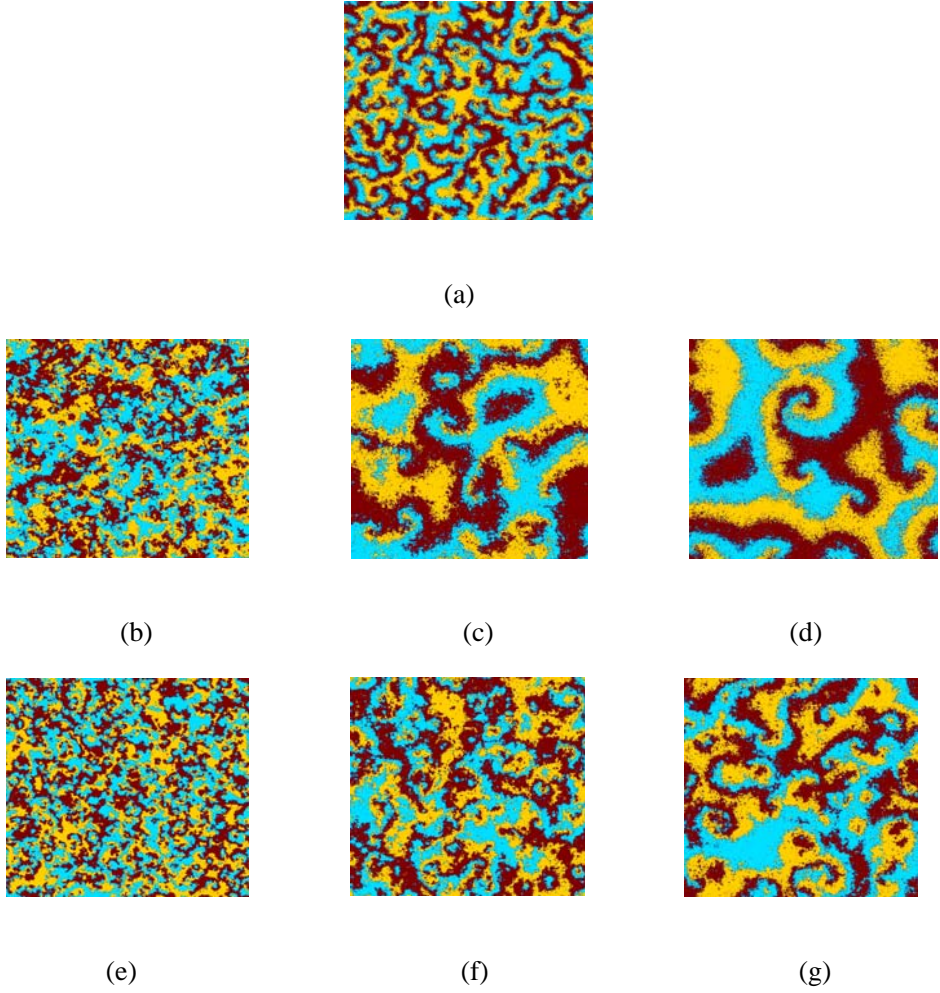

Fig.4. Snapshots of patterns. The size of square lattice is $500 \times 500$. For each realization, the simulation consists of $10^4$ time steps. The maximal step length $l_{max}$ of Levy flight equals 50.

(a) $e = 0.2, p_0 = 1.0$.

(b) $h = -1.5, e = 0.2, p_0 = 0.9$. (c) $h = -1.5, e = 0.2, p_0 = 0.5$.
(d) $h = -1.5, e = 0.2, p_0 = 0.0$. (e) $h = -2.0, e = 0.2, p_0 = 0.9$.
(f) $h = -2.0, e = 0.2, p_0 = 0.5$. (g) $h = -2.0, e = 0.2, p_0 = 0.0$.

bigger patches grow in size gradually. On the condition that $p_0 = 1.0 \to 0.0$, $h = -3.0 \to -1.0$, and $e$ unchanged, the patterns vary from spiral wave to fragmentized patches, and then to spiral wave again with bigger size of patches. In order to see the change of spatiotemporal patterns clearly, we have done the stochastic lattice simulations in the lattice with $500 \times 500$ sites. The results demonstrate the same phenomena. The effects of lone-range interaction on the final spatiotemporal patterns are interesting and are needed for further investigation.

## 4. Concluding remarks

In this paper, we have mainly studied the interplay between NNI and LRI of Levy flight in spatial rock–paper–scissors game. Our aim is to find the effects of the long-range-interactions on the biodiversity in ecosystem. The coexistence condition of three species, such as the critical curve with the change of parameter $p_0$, power-law exponent $h$, and maximal step length $l_{max}$ of Levy flight are presented. The curves also show the change of the area representing stable coexistence of three species in the parameter space. When the maximal step length $l_{max}$ approximately equals half of lattice's length, the area of stable coexistence of three species tend to a minimal limit. All the simulation results show that the probability $p_0$, maximal step length $l_{max}$, and power-law exponent $h$ of Levy flight make distinct impacts on the stability and the forms of the spatiotemporal patterns. Long-range-interactions with Levy flight are common in the real ecosystems. It needs further investigation to reveal the effects and the corresponding mechanisms of Levy flight on the biodiversity.

## Acknowledgments

This work was partially supported by the NSFC under the Grant Nos. 61174150 and 60974084, the Program for New Century Excellent Talents in University of Ministry of Education of China (No. NCET-09-0228), fundamental research funds for the Central Universities of Beijing Normal University, and HSCC of Beijing Normal University.